# Technological Requirements for Videoconferencing Judicial Hearings: Enhancing the Credibility and Reliability of Remote Testimonies

Jorge Alberto Araujo

Tribunal Regional do Trabalho da 4ª Região (TRT4)[1], Porto Alegre, Brazil

jaaraujo@trt4.jus.br

## Abstract

This paper analyzes the technological requirements necessary to enhance the credibility and reliability of judicial hearings conducted via videoconference, from the internal perspective of the judiciary. Drawing on the practical experience of a judge who conducts daily hearings, this study identifies limitations in current platforms for verifying the authenticity of testimonies and proposes tailored functionalities for the judicial context. Recognizing that remote hearings represent a convenience for the parties, without replacing the option of in-person attendance, the article suggests implementing features such as eye tracking, environment verification, and blocking of parallel applications, in addition to improvements in transmission quality. The study concludes that developing specific modules for witnesses—focusing on security and monitoring—can significantly contribute to equalizing the credibility between remote and in-person hearings, thus expanding access to justice without compromising procedural reliability.

---

[1] The views expressed in this article are those of the author and do not necessarily reflect the official position of the TRT4.

## 1 Introduction

The conduct of judicial hearings through videoconferencing represents one of the most significant technological transformations in the contemporary justice system. What initially emerged as an emergency solution to ensure the continuity of judicial services during periods of mobility restrictions, such as the Covid-19 pandemic, which occurred primarily between 2020 and 2022, quickly consolidated as an important complementary practice, offering substantial benefits in terms of accessibility, procedural economy, and celerity[2]. In Brazil, this hearing modality received legal backing through several resolutions of the National Council of Justice (Conselho Nacional de Justiça - CNJ), culminating in CNJ Resolution No. 354/2020 [1], which legally equated remote hearings with in-person hearings for all legal purposes[3].

Remote hearings offer a convenient alternative for parties and other participants, helping them avoid the often costly and time-consuming travel to court [2]. However, the option of in-person attendance remains available, ensuring that individuals with technological limitations or other challenges can still exercise their right to participate fully. This coexistence of formats reinforces the principle of access to justice by accommodating the diverse realities of those subject to the court's jurisdiction [3].

Despite the evident benefits, remote hearings present significant challenges regarding the guarantee of credibility and reliability of testimonies [4]. One of the main problems identified in daily judicial practice refers to the possibility of participants, especially witnesses, using external resources during their testimonies, such as consulting notes, communicating with third parties, or accessing unauthorized information [1] [5]. This vulnerability compromises one of the fundamental principles of the judicial process: the authenticity and spontaneity of witness testimonies [4].

---

[2] The ordinance that officially ended Brazil's Public Health Emergency of National Importance due to COVID-19 was signed on April 22, 2022.

[3] The National Council of Justice (Conselho Nacional de Justiça, CNJ) is a constitutionally-created body (Constitutional Amendment No. 45/2004) charged with overseeing administrative and financial performance of Brazil's courts, ensuring transparency, efficiency, and adherence to judicial ethics across the federal and state judiciaries.

Another critical aspect relates to the difficulty of observing participants' behavior [6]. In in-person hearings, judges, lawyers, and other legal operators can observe bodily reactions and facial expressions that help interpret or contextualize testimonies. In the videoconference environment, these observations become limited by transmission quality, restricted framing, and the possibility of environment manipulation by the participant [7].

Based on practical judicial experience, this article argues that remote hearings are not a temporary fix nor an inferior alternative to in-person hearings. Instead, they represent a necessary and irreversible evolution of the judicial system [2]. However, for them to achieve full equivalence in terms of credibility and evidentiary effectiveness, it is essential that the technological tools employed be enhanced with specific functionalities for the judicial context. After all, the platforms currently used were designed for collaborative settings—such as team presentations or alignment meetings—and not for the adversarial environment inherent to the process [5], in which parties with opposing interests dispute conflicting versions of facts[4].

The central objective of this study is to identify and propose technological requirements that videoconferencing platforms should incorporate to enhance the credibility and reliability of judicial hearings via videoconference. Specifically, it seeks to: (1) Analyze the limitations of current tools in the judicial context [8]. (2) Identify monitoring and verification technologies applicable to the remote hearing environment. (3)

---

[4] The videoconferencing platforms most widely used in the Brazilian judiciary—such as Zoom, Microsoft Teams, and Google Meet—were originally developed for corporate and educational environments, where collaborative interactions prevail. The judicial setting, by its nature, is characterized by adversarial proceedings and the need for specific procedural safeguards, requiring features that generic platforms do not adequately provide.

Propose specific functionalities for controlling the environment and behavior of witnesses. (4) Suggest protocols that ensure balance between procedural security and fundamental rights of participants.

The relevance of this study stems from the growing adoption of remote hearings in the judicial system, in Brazil and worldwide—a trend that persists even after the emergency period of the pandemic [9]. By proposing specific technological enhancements for the courtroom environment, from the perspective of someone who uses these tools daily in conducting hearings, this work contributes to the improvement of digital platforms [10], expanding their operational advantages without compromising process reliability.

This study uses specific concepts and terminology from both the legal and technological fields. To facilitate understanding of the technical terms used throughout the text, especially those related to monitoring and verification technologies in videoconference environments, a glossary has been included at the end of the article (before the bibliographic references). Its consultation is recommended for greater clarity in understanding the proposed solutions.

## 2 Methodology

This study adopted a qualitative and exploratory approach, combining literature review, document analysis, and the proposition of technological solutions, grounded in the practical experience of a judge who conducts hearings daily. The methodology was structured in four main stages, described below.

In the first stage, a comprehensive literature review was conducted on judicial hearings via videoconference, credibility of testimonies in digital environments, non-verbal behavior in judicial contexts, and applicable monitoring technologies. Scientific articles, technical studies, institutional reports, and specialized publications were consulted, prioritizing sources from the last five years, but including older seminal references when relevant to the theoretical foundation.

The second stage consisted of document analysis of the rules and regulations governing remote hearings in Brazil, with special focus on the resolutions of the National Council of Justice (CNJ) and normative acts of courts [1]. This analysis sought to understand the legal and institutional framework that supports the conduct of remote hearings, identifying requirements, limitations, and guidelines established by regulatory bodies.

In the third stage, an analysis of videoconferencing tools currently used in the Brazilian judicial system was conducted, with emphasis on Zoom, a platform officially adopted by various courts, including the Superior Labor Court (Tribunal Superior do Trabalho - TST) [11]. This analysis contemplated the identification and evaluation of available resources, technical limitations, and vulnerabilities in relation to the specific needs of the judicial context. Technical manuals, institutional tutorials, and official platform documentation were consulted, in addition to practical experience accumulated in conducting remote hearings over several years.

Finally, the fourth stage involved the elaboration of proposals for technological requirements to enhance videoconferencing tools in the judicial context. This elaboration was based on findings from previous stages, practical judicial experience, and identification of promising technologies already applied in other contexts, such as proctoring systems for online exams [12]. The proposals were developed considering criteria of technical feasibility, adequacy to the judicial context, respect for fundamental rights of participants, and potential contribution to the credibility of remote hearings.

It is important to note that this study did not conduct empirical tests of the proposed technologies, being a theoretical-propositional research grounded in practical experience.

The technical validation of the presented proposals constitutes a subsequent stage, which would require implementation and evaluation in controlled environments, which is suggested as a direction for future research.

## 3 Context and Limitations of Remote hearings

### 3.1 Remote hearings in the Brazilian judiciary: Evolution and Impacts

The conduct of remote hearings in the Brazilian judicial system is not a recent innovation, although it has intensified and consolidated in recent years. The initial milestone of the regulation of this modality dates back to 2010, when the National Council of Justice (CNJ) issued Resolution 105/10, which provided for the conduct of interrogations and witness inquiries by videoconference [13]. However, it was from 2020, driven by the need for continuity of judicial services during the COVID-19 pandemic, that the use of videoconferencing tools for judicial hearings expanded significantly.

In response to the new reality, the CNJ published Resolution No. 354/2020, in November 2020, which comprehensively regulated the conduct of hearings and sessions via videoconference and other remote means. This resolution represents a fundamental milestone by legally equating remote hearings to in-person ones for all legal purposes [1], ensuring the publicity of acts performed and the procedural prerogatives of lawyers, members of the Public Ministry, public defenders, parties, and witnesses.

The norm establishes important definitions, distinguishing two fundamental concepts:

- "Videoconference session": remote communication conducted in judicial unit environments.
- "Remote session": hearings and sessions conducted from a physical environment outside the courthouse.

Additionally, the resolution determines that hearings be recorded and the audiovisual file be attached to the case records or made available in an official media repository indicated by the CNJ (PJe Mídias) or by the court [1].

The adoption of remote hearings in Brazil followed a global trend of digitization of judicial services, but presented its own characteristics, such as the rapid adaptation of courts and the creation of specific regulations.

According to CNJ data, during the period from 2020 to 2022, millions of remote hearings [11] were conducted throughout the national territory, demonstrating the consolidation of this modality as an integral part of the Brazilian judicial system.

The identified benefits include the reduction of travel costs, the decrease in case processing time, the possibility of participation by people in different locations, and the continuity of judicial services even in adverse situations. In my experience as a judge, I have observed that remote hearings have been particularly beneficial for witnesses residing in locations distant from the court headquarters, for parties with mobility difficulties, and for lawyers who work in multiple jurisdictions.

However, studies such as the one by the **Surveillance Technology Oversight Project** point to significant challenges, such as the loss of important non-verbal information, attorney-client confidentiality issues, and the potential dehumanization of the judicial process [23]. These challenges are perceived in the daily practice of judicial courts and demand specific solutions, such as those proposed in this article.

### 3.2 Credibility of Testimonies in Remote Hearings: Limitations

The credibility of testimonies given in judicial hearings constitutes a fundamental element for the formation of the judge's conviction [4] and, consequently, for the quality of judicial service. In the context of remote hearings, this credibility faces specific challenges that deserve in-depth analysis.

The study released by the Surveillance Technology Oversight Project in 2020—previously cited—identified that digital trials can harm the defense in various aspects, placing defendants at a disadvantage. Among the problems pointed out, the difficulty of observing bodily expressions, facial manifestations, and other behavioral nuances [6] that are essential for contextualizing testimonies stands out.

Research in the field of testimony psychology suggests that the evaluation of testimony credibility relies on two essential components:

1. The verbal content of the statements.
2. A series of non-verbal elements, such as eye contact, hesitations, and body posture [14].

In a videoconference environment, many of these non-verbal elements can be lost or distorted due to technical limitations, such as image quality, restricted framing, and transmission delays.

Another critical aspect refers to the possibility of external interference during testimony. In in-person hearings, the controlled environment of the hearing room minimizes the possibility of consulting unauthorized materials or communicating with third parties. In the videoconference environment, however, the participant can easily access documents, receive guidance through messages, or have other people present in the same environment, outside the camera's field of view [5].

A study conducted by Diamond et al. (2010) in Cook County, Illinois (USA), revealed that defendants who appear virtually [15] at bail hearings are typically disadvantaged, with bail being set on average 51% higher compared to defendants who appear physically before the judge. This data suggests that technological mediation can significantly affect the perception of credibility and, consequently, judicial decisions.

In Brazil, although comprehensive quantitative studies on the impact of remote hearings on the credibility of testimonies are still lacking, daily judicial experience reveals recurring concerns. Among them, the difficulty in evaluating the sincerity and spontaneity of statements, the impossibility of ensuring that the witness is not being guided [16], and the loss of relevant non-verbal elements stand out.

In the daily routine of courts, for example, it is common for lawyers to complain about the behavior of witnesses presented by the opposing party, frequently accused of averting their gaze from the camera, which would suggest possible consultation of notes or communication with third parties.

In my experience as a judge, I have observed that these limitations frequently generate problematic situations during hearings. On several occasions, it was possible to perceive that witnesses were clearly consulting materials outside the camera's field of view, or being instructed by third parties present in the same environment, or receiving guidance through instant messaging apps such as WhatsApp [12]. The technical impossibility of verifying and controlling these situations compromises the credibility of testimonies and, consequently, the quality of judicial service.

### 3.3 Videoconferencing Tools in the Judiciary: Resources and Limitations

### 3.3.1 Overview of Platforms Used in the judiciary

The Brazilian judicial system has adopted various videoconferencing platforms to conduct remote hearings, with Zoom predominating [17] as the official solution in many courts. The selection of platforms by judicial bodies was based on criteria such as stability, recording capacity, ease of use, and security features, as well as considerations regarding costs and scalability.

The Superior Labor Court (TST) mandated that, starting in May 2021, all hearings and trial sessions within the labor judiciary be conducted exclusively via the Zoom platform[5] [17]. This standardization aimed to unify procedures and facilitate access for lawyers and parties operating across different courts and tribunals.

In addition to Zoom, other platforms such as Microsoft Teams, Google Meet, and Cisco Webex are used to a lesser extent by some courts or in specific situations. While this diversity of solutions offers flexibility, it also poses challenges related to interoperability and procedural standardization [8].

The adoption of these commercial platforms, originally developed for corporate and educational contexts, represents a pragmatic adaptation to the judiciary's emergency needs. However, as detailed in the following sections, these tools were not designed specifically for the judicial context, presenting significant limitations [18] in meeting the specific requirements for the credibility and reliability of testimonies.

---

[5] *Adoption of Zoom by the Brazilian Labor Judiciary.* The platform Zoom was officially adopted as the standard videoconferencing system for all hearings and trial sessions within the Brazilian labor courts, following an institutional effort to unify procedures nationwide.

### 3.3.2 Existing Resources and Functionalities

The videoconferencing platforms used in the judiciary offer a set of features that, although not developed specifically for the judicial context, have been adapted to meet the basic needs of remote hearings. A detailed analysis of Zoom, as the predominant platform, reveals relevant functionalities grouped into four main categories: security and control; recording and documentation; collaboration; and accessibility.

In terms of security and control, Zoom provides features such as a waiting room, which allows the host (judge or court clerk) to control who enters the hearing; user authentication, requiring participant identification; communication encryption; screen-sharing control; meeting lock after the session begins; participant removal; remote deactivation of audio and video; and access passwords. These features provide a basic level of security, enabling access control and participant management during the hearing.

Regarding recording and documentation, the platform supports cloud or local recording of hearings, with the option for automatic transcription; participant registration; and export of recordings in standard formats. These resources are essential for the official documentation of procedural acts, as required by CNJ Resolution No. 354/2020.

Collaboration features include simultaneous breakout rooms, which enable private conversations between lawyers and clients; document sharing for presenting evidence; public and direct (private) messaging; and shared annotations. These functionalities facilitate interaction among participants and the presentation of relevant materials during the hearing.

Finally, in terms of accessibility, Zoom offers access across multiple devices (computers, tablets, smartphones); a relatively intuitive interface; captioning options; and quality adjustments based on available connectivity. These features contribute to expanding access to remote hearings, reducing technological barriers.

Other platforms, such as Microsoft Teams, Google Meet, and Cisco Webex, offer similar sets of functionalities with specific variations. For example, Microsoft Teams provides better integration with the Office ecosystem, while Webex stands out for additional security features tailored to government institutions.

### 3.3.3 Limitations and Vulnerabilities

Despite the available resources, current videoconferencing platforms present significant limitations [18] when analyzed from the perspective of the specific needs of judicial hearings, particularly in ensuring the credibility and reliability of testimonies.

One of the most critical limitations is the absence of eye-tracking capabilities. None of the analyzed platforms offer tools to verify where a participant is looking during the hearing [6], making it impossible to detect whether a witness is consulting notes, receiving written instructions, or reading prepared responses. This limitation significantly undermines the spontaneity and authenticity of testimonies, which are essential for their evidentiary value.

Another significant vulnerability is the lack of blocking parallel applications. During a videoconference hearing, participants can easily access other programs, browse the internet, exchange messages [5], or consult unauthorized documents without detection by other participants or the system. This possibility starkly contrasts with the controlled environment of in-person hearings, where such actions would be immediately noticeable.

The use of virtual backgrounds also represents a relevant limitation. All analyzed platforms allow the use of virtual backgrounds or environment blurring [7], features that, while useful for privacy in general contexts, can be problematic in judicial hearings by concealing the participant's actual environment. This functionality may prevent verification of who else is present or what materials are available for consultation.

Equally problematic is the lack of environment verification [16]. None of the platforms require or facilitate a comprehensive check of the physical space where the participant is located, making it impossible to ensure that a witness is alone or free from external influences during their testimony.

The absence of a dedicated 'witness mode' [18] constitutes another significant gap. The platforms provide the same functionalities to all participants, without accounting for the distinct roles and heightened security requirements for witnesses, who should be subject to stricter controls to ensure the reliability of their testimonies.

Finally, the lack of environmental sound monitoring [12] represents an additional vulnerability. Current platforms do not detect or alert to external communications or suspicious noises that could indicate the presence of others guiding the witness's testimony.

Taken together, these limitations reveal a significant gap between the specific needs of the judicial context and the functionalities available in commercial videoconferencing platforms. This gap justifies the development of tailored technological solutions or the adaptation of existing platforms to meet the essential requirements for credibility and procedural reliability in judicial videoconference hearings.

In my experience as a judge, I have observed that these limitations frequently lead to problematic situations during hearings. On numerous occasions, it was evident that witnesses were consulting materials outside the camera's field of view, being instructed by third parties present in the same environment, or receiving guidance through instant messaging applications such as WhatsApp. The lack of technical means to verify or control these situations undermines the credibility of testimonies and, consequently, the quality of judicial service.

## 4. Technological Proposals and Implications

### 4.1 Proposed Technological Functionalities

#### 4.1.1 Specific Module for Witnesses

The analysis of the limitations of current platforms evidences the need for a specific module for witnesses, with dedicated functionalities to ensure the credibility and reliability of testimonies. This module should incorporate eye and attention monitoring technologies, resource and application blocking, and environmental check, as explained in the following subsections.

Eye and attention monitoring represents a core element of this module. The proposal draws inspiration from technologies already adopted in high-stakes settings—such as university admissions and visa evaluations—that employ eye-tracking to ensure integrity, exemplified by the Duolingo English Test [12]. In the judicial context, this functionality would track the direction of the witness's gaze to verify whether their attention remains focused on the screen or frequently shifts elsewhere. Using the device's camera and image processing algorithms, the system would map eye movements and generate alerts when suspicious or irregular visual patterns are detected.

Complementarily, the implementation of attention deviation detection is suggested, which continuously monitors the witness's level of attention through indicators such as head position and gaze direction. The system to the court operator when the witness appears not to be concentrating on the hearing for prolonged periods.

This monitoring should also prevent the use of any types of filters or additional tools, especially those that allow, for example, simulated eye contact. Such resources, although useful in social or corporate contexts, can compromise the authenticity of communication in a judicial context [4].

In the scope of resource and application blocking, the implementation of an exclusive full-screen mode is proposed that prevents the witness from minimizing the application or accessing other programs during testimony. This mode should be automatically activated at the beginning of the testimony and can only be deactivated by the host [18] (judge or secretary) or at the end of the session.

Additionally, the detection of running applications is suggested, which verifies and reports which programs are active on the witness's device before and during testimony. This functionality can identify potentially problematic applications, such as messaging programs, secondary browsers, or remote access tools.

Temporary notification blocking is also recommended, suspending alerts, messages, and other interruptions during testimony. This function minimizes distractions and reduces the possibility of unauthorized external communication.

This module could even restrict the use of mobile devices through existing control apps, or implement another way to prevent communication via cell phone in messaging services such as WhatsApp. One possibility would be integration with parental or business control applications [7] that can temporarily restrict the use of certain applications on mobile devices.

Complementing these measures, network connection monitoring is proposed, which detects and reports parallel communications via the internet during testimony. This functionality can identify attempts to access websites, exchange messages, or other forms of external communication.

Environment verification constitutes another essential component of the witness module. Mandatory 360° scanning is proposed, requiring witnesses to rotate their camera to verify their environment before testifying. This procedure allows the judge and parties to confirm that there are no other people or unauthorized materials at the location [16].

Human presence detection configures a complementary functionality, based on computer vision and machine learning algorithms [14] capable of identifying the existence of other people in the environment, even if partially visible or outside the main framing of the camera. This technology can analyze reflections, shadows, sounds, and other indirect indicators of presence.

Continuous acoustic monitoring is also proposed, through analysis of the sound environment with the objective of detecting voices, whispers, or any noises that may indicate communication with third parties. Audio processing algorithms allow filtering normal ambient sounds and highlighting those potentially relevant to the reliability of the testimony.

Finally, the implementation of periodic random verification is suggested, with unexpected requests for new environment scanning during testimony. Such functionality aims to hinder premeditated attempts at fraud, since the witness will not know when they will be prompted to display the location again [9] where they are.

### 4.1.2 Protocols for Display and Control of Participants

In addition to technological functionalities, it is fundamental to establish clear protocols for the display and control of participants during remote hearings. These protocols should be implemented as technical requirements of the platforms, ensuring their consistent application.

An essential requirement is that participants, especially witnesses and parties, display at least their face and part of their torso [4] during the entire hearing. This broader display allows the observation of body language and important non-verbal signals for contextualizing testimony. The platform should implement an alarm or notification system when the participant does not remain adequately visible according to the established standard.

Another fundamental aspect is centralized control by the host (judge or secretary) over the audio and video functionalities of participants. Witnesses and parties should not have the possibility of turning off their cameras or microphones on their own during testimony. This is an exclusive prerogative of the host [7]. This restriction should be clearly informed at the beginning of the hearing and be part of the initial participation declaration.

Additionally, witnesses and parties should not have access to the platform's chat or meeting update tools [5] such as, for example, the Zoom Companion, which could be used for unauthorized communication or manipulation of hearing settings. The platform should allow the configuration of different levels of access according to the participant's role, automatically restricting these functionalities for witnesses.

The implementation of these protocols requires a formal initial declaration, where the participant identifies themselves, confirms their participation in the hearing, and acknowledges the technical limitations imposed to ensure the reliability of the procedure. This declaration should be recorded and stored [6] as part of the official record of the hearing.

An alternative, perhaps even more viable, to this initial formal declaration, would be the presentation of such limitations in a notice (pop-up) in the tool itself, as currently occurs when the system notifies users about recording at the start of a hearing.[6]

### 4.1.3 Resources for High-Quality Visualization and Recording

To approximate remote hearings to the informational richness of in-person proceedings, it is essential to implement advanced visualization and recording tools, focusing on transmission quality without relying on automated behavioral analysis.

In terms of enhanced visualization, the use of mosaic mode with active speaker highlighting is proposed. This feature offers two key advantages:

1. It enables the simultaneous display of all participants [7].
2. It automatically highlights the current speaker.

This resource facilitates contextual observation of participants, allowing judges and other legal professionals to perceive reactions and interactions that would naturally occur in an in-person setting.

Another important feature is disabling virtual backgrounds. It enables judges to see participants without visual filters that may obscure key elements of their physical surroundings.This functionality can be activated at specific moments—respecting the witness's privacy while ensuring the integrity of the hearing when needed. In the experience of the judicial practice, the use of virtual backgrounds frequently hinders the perception of essential contextual factors, such as the possible presence of third parties [5] or consultation of unauthorized materials.

---

[6] **Interface-Based Consent Mechanisms.** Although the requirement for a pop-up consent notice is not explicitly stated in the regulation, it aligns with the procedural safeguards established by Brazilian judicial norms. Article 7, item IV of *Conselho Nacional de Justiça*, *Resolução* No. 354/2020, mandates that videoconference hearings be recorded and that the audiovisual file be added to the case records or made available in an official repository. This provision supports the implementation of a pop-up or interface-based notice to inform participants and formalize their acknowledgment of the recording.

High-definition video transmission is an essential technical requirement [4]. It enables accurate observation of facial expressions and body language, supporting the contextual evaluation of testimonies. Unlike automated behavioral analysis systems—which may raise ethical and legal concerns—high-definition video merely enhances the technical conditions under which legal professionals exercise their interpretive judgment based on experience and legal expertise.

Regarding recording and documentation, the adoption of multidimensional recording is suggested, with individualized storage of audio and video of each participant [15]. This format allows clear review of testimonies, without visual interference from other interlocutors—especially useful in processes with multiple witnesses or extensive hearings.

As a complementary resource, it is recommended to maintain mosaic recording, which preserves the interactional context of the hearing. This approach records simultaneous reactions of participants, offering the judge a richer view of the dynamics present in the session.

Manual time markers also represent a valuable functionality. They allow the judge or secretary to insert markings during the hearing, with brief descriptions (e.g., "beginning of witness testimony," "contradiction about work hours," "clarification about hierarchy"), facilitating navigation and location of relevant excerpts [8] during case analysis.

In the automated documentation plan, in addition to the transcriptions already discussed, the implementation of intelligent indexing of audiovisual content is proposed, with keyword searches in recordings [12]. Such functionality would combine speech recognition and textual indexing, allowing legal operators to quickly locate specific segments of testimonies, without the need to watch the entire recording.

It should be emphasized that all these functionalities should be conceived as support tools, and not as a replacement for human analysis. It is up to legal professionals, based on established legal criteria, to interpret testimonies. The role of technology, in this context, is to expand the perceptive and analytical capabilities of legal operators, without usurping their technical or evaluative judgment.

In this same sense, it is essential to ensure that all generated content—including recordings in different formats, time markers, and indexing—is fully accessible to all legal operators who act in the process, especially to the parties and their attorneys. It is not enough to make available only one recording modality to the public or to the court: it is necessary to ensure full access to all forms of registration as a condition for the effective exercise of the adversarial principle and for the comprehensive appreciation of the evidence produced.

### 4.1.4 Transcription Modalities and Their Purposes

The documentation of remote hearings should contemplate three complementary modalities of transcription, adaptable to different procedural purposes:

**Verbatim transcription**: Faithfully records everything that was said by participants, including hesitations, repetitions, pauses, interruptions, language vices, and emotional expressions. It is especially useful for credibility analysis, detection of discursive inconsistencies, and evaluation of the emotional context of the testimony.

**Edited transcription**: Presents the substantive content in a cleaner form, with basic language corrections, suppression of vices, and reorganization of interrupted phrases. It facilitates reading, understanding of the main arguments, and identification of key points of the narrative.

**Intelligent transcription with markers**: Combines the clarity of the edited version with specific annotations about relevant pauses, significant hesitations, and observed non-verbal behaviors. It allows fluid reading, preserving contextual elements useful for credibility evaluation.

The videoconference system should allow prior or posterior configuration of the desired modality, enable switching between views, and offer specific search and filtering resources. The verbatim version should always be preserved as the primary record, ensuring procedural reliability and reproducibility of evidence.

Flexibility in transcription modalities represents an important advance in the documentation of remote hearings, allowing different aspects of testimonies to be highlighted according to the specific needs of each procedural phase. In forensic practice, this diversity has proven especially useful in complex cases, with multiple witnesses or extensive hearings.

### 4.2 Technical Feasibility of the Proposed Solutions

The functionalities and protocols proposed in this study represent a significant advancement over the videoconferencing platforms currently used in the judicial system. However, it is essential to assess their technical feasibility, considering the current state of technology, infrastructure requirements, and potential implementation challenges.

Eye monitoring and attention detection, central components of the module designed for witnesses, are based on well-established technologies applied in various contexts. Eye-tracking systems have been used for years in usability research, marketing, and, more recently, in proctoring platforms for online exams, as already applied in high-stakes online exams, such as the Duolingo English Test and other remote proctoring systems used in university admissions [12]. These technologies rely on conventional cameras combined with image-processing algorithms to map gaze direction, requiring no specialized hardware. Their implementation in videoconferencing platforms is technically feasible [7], though it necessitates adaptations to ensure reliable performance across diverse devices and lighting conditions.

Blocking parallel resources and applications poses more significant technical challenges [12], particularly in uncontrolled environments. Educational proctoring solutions already implement similar functionalities, but they often require installing specific software with elevated operating system permissions. In the judicial context, solutions balancing effectiveness and accessibility would need to be developed, possibly combining browser-based approaches with complementary extensions or applications. An alternative would be the creation of secure digital environments (sandboxes) that isolate the videoconferencing application from other system resources.

Environment verification, including mandatory 360° scanning and human presence detection, is technically feasible with cameras available in common devices. Computer vision algorithms and artificial intelligence for person detection are widely used in security systems and could be adapted for this context. The primary challenge lies in the variability of domestic environments and the need to distinguish between authorized and unauthorized presences.

Enhanced visualization and multidimensional recording resources are technically viable, with some already implemented in specialized systems. For instance, separate recording of each participant is a feature available in certain webinar platforms. The main innovation proposed here involves integration with indexing systems and consistent transmission quality, which would require specific optimizations for the judicial context.

Security protocols and centralized host control are relatively straightforward to implement from a technical perspective, as they primarily involve configuring permissions and user interfaces. Many platforms already offer granular permission controls that could be adapted to meet the specific requirements of remote hearings.

In summary, most of the proposed functionalities are technically feasible with current technology, albeit at varying levels of maturity. Some could be implemented in the short term as extensions of existing platforms, while others would demand more extensive development. A gradual and modular approach, prioritizing functionalities with high impact and lower technical complexity, would be advisable to facilitate the progressive adoption of these innovations[7] [2].

### 4.3 Ethical and Legal Considerations

The implementation of the functionalities and protocols proposed in this study raises important ethical and legal considerations that must be carefully analyzed to ensure that the technological advancement of remote hearings aligns with the fundamental principles of a democratic rule of law.

The first ethical dimension to consider is the balance between procedural security and access to justice. Technologies such as eye monitoring, environment verification, facial recognition, and behavioral analysis, while valuable for procedural reliability, represent intensive forms of surveillance that may be perceived as invasive. Their implementation must be guided by the principles of proportionality and necessity, limiting access to only what is strictly required for the legitimate purposes of the judicial process.

In this regard, transparency and informed consent are indispensable ethical prerequisites. All participants must be clearly informed about what data is collected, how it is processed and stored, and for what specific purposes. Consent should be obtained freely, using accessible language and ensuring that participants genuinely understand the implications of their decision. However, in judicial settings, the notion of voluntary consent becomes problematic, as declining to consent may carry implicit or explicit procedural consequences—particularly for vulnerable individuals or those in asymmetrical positions within the case. This underscores the need for safeguards that ensure consent is not merely formal, but substantively free and fair.

Compliance with data protection legislation—particularly Brazil's General Data Protection Law (LGPD, Law No. 13.709/2018)[8]—is an unavoidable legal imperative. The processing of sensitive data by proposed technologies must strictly adhere to the principles of purpose, necessity, and transparency established by this framework. The proposed technologies involve processing sensitive personal data, including biometric information (e.g., facial recognition, eye movement patterns) and potentially data that could potentially reveal psychological or emotional traits. Such processing must strictly adhere to principles of purpose limitation, data minimization, necessity, accuracy, transparency, security, accountability, accessibility, and non-discrimination.

---

[7] *Phased Implementation of Legal Technologies*. Incremental deployment allows adaptation to judicial infrastructure, minimizing resistance and ensuring better user integration.
[8] See Glossary.

Procedural safeguards and participants' fundamental rights represent another critical area of concern. The right to a fair defense and due process requires that all parties have equal access to the implemented technologies and an adequate understanding of their operation. The principle of equality of arms could be undermined if only one party possesses sufficient technical knowledge to interpret or challenge the results of these analyses. Additionally, the right against self-incrimination could be affected by technologies that inadvertently analyze revealing behavioral traits.

Technological accessibility also carries significant ethical implications. The adoption of advanced technologies must not create additional barriers to access to justice. Participants with technological limitations—whether due to economic, educational, or disability-related factors—must be provided with suitable alternatives to avoid procedural disadvantages. This underscores the importance of maintaining the option for in-person court attendance, as emphasized earlier in this article.

Finally, the interpretation of results presents a significant ethical-legal challenge. Behavioral analyses must be treated as supplementary tools, never as substitutes for human judgment. The risk of attributing excessive evidentiary weight to technological indicators, known as "automation bias," could compromise the quality of judicial service. Clear guidelines on the weight assigned to these analyses and adequate training for legal professionals to interpret them critically are essential.

## 4.4 Limitations of the Study and Future Research

This study presents several limitations that should be acknowledged and addressed in future research. The theoretical-propositional nature of the work, while allowing for comprehensive exploration of potential solutions, and does not empirically validate the proposed technologies [10]. The functionalities suggested were not subjected to practical testing in real judicial environments, which limits the ability to assess their actual effectiveness, usability, and acceptance by different stakeholders.

Another significant limitation relates to the diversity of judicial contexts. The Brazilian judicial system is heterogeneous in terms of technological infrastructure [11], procedural practices, and digital familiarity of operators. The proposals presented may require substantial adaptations for implementation in different courts and instances, considering their specificities and limitations.

The cost-benefit analysis represents a dimension not addressed in depth [2] in this study. The implementation of the proposed technologies would involve significant investments in development, infrastructure, training, and technical support. A rigorous evaluation of the relationship between these costs and the expected benefits in terms of jurisdictional quality would be essential to inform implementation decisions.

The perspective of users, particularly witnesses and parties not accustomed to the judicial environment, remains an unexplored aspect. The acceptability of the proposed technologies, their usability, and their impact on the subjective experience of participants are aspects that require specific investigation [19], preferably with participatory methodologies that incorporate the perceptions and concerns of the various actors involved.

These limitations point to several directions for future research. Controlled experimental studies could test the effectiveness of the proposed functionalities [20] in simulated judicial hearings, evaluating both their technical performance and their impact on the perception of credibility. Comparative research could analyze different technological configurations, identifying the optimal set of functionalities that enhance credibility while reducing perceived intrusiveness.

Interdisciplinary investigations involving law, psychology, computer science, and applied ethics [21] would be valuable to deepen the understanding of the multidimensional implications of the proposed technologies. Particularly relevant would be the study of potential biases in behavioral analysis and the development of methods for their mitigation.

Field research in courts that already implement preliminary versions of these technologies could provide valuable insights [3] into practical implementation challenges, acceptance by legal operators, and impact on procedural dynamics. Longitudinal studies could evaluate how increasing familiarity with these technologies affects their effectiveness and acceptance over time.

Finally, investigations into the economic and judicial management dimension would be essential to inform public policies for technological modernization of the judiciary. Cost-effectiveness analyses, gradual implementation models, and large-scale training strategies represent promising areas for applied research that could facilitateto practical implementation in daily judicial routines.

## 5 Implications and Conclusions

This study analyzed the technological requirements necessary to enhance the credibility and reliability of judicial hearings conducted via videoconference, identifying limitations of current platforms and proposing specific functionalities for the judicial context. The research was grounded in an internal judicial perspective, based on the practical experience of a judge who conducts hearings daily, and recognizing that remote hearings, although representing a necessary evolution of the judicial system, should coexist with the in-person option to ensure full accessibility [9].

The analysis of videoconferencing tools currently used in the Brazilian judiciary, with emphasis on Zoom as the predominant platform, revealed significant limitations regarding the specific needs of the judicial context

[18]. The absence of eye monitoring, the lack of blocking of parallel applications, the possibility of using virtual backgrounds that hide the real environment, the lack of environment verification, and the absence of specific resources for centralized control represent vulnerabilities that compromise the credibility of videoconference testimonies.

In response to these limitations, specific functionalities were proposed, organized into three main categories. The specific module for witnesses incorporates eye and attention monitoring technologies, blocking access to other applications and system resources, and environment verification, aiming to ensure that the witness does not consult unauthorized materials or receive external guidance during testimony. The protocols for display and control of participants establish requirements for the visual presentation of deponents and restrictions on access to functionalities that could compromise the reliability of the process. The resources for high-quality visualization and recording [15] propose improvements in audiovisual transmission and recording, allowing better observation and documentation of testimonies.

The discussion on the technical feasibility of the proposed solutions indicated that most functionalities are technically feasible with current technology, although at different levels of maturity. Some could be implemented in the short term as extensions of existing platforms, while others would require more substantial development. The analysis of potential impacts on the credibility of remote hearings suggested that the proposed functionalities could significantly mitigate the disadvantages of the videoconference environment, narrowing the informational gap between remote and in-person hearings.

The ethical and legal considerations highlighted the need to balance procedural security and privacy, ensure transparency and informed consent,comply with data protection legislation, preserve procedural guarantees and fundamental rights, and promote technological accessibility. In this context, maintaining the option for in-person attendance emerges as a fundamental guarantee to ensure access to justice for all citizens [8], regardless of their technological limitations.

The limitations of the study, including its theoretical-propositional nature, the diversity of judicial contexts, the absence of in-depth cost-benefit analysis, and the limited consideration of the user perspective, point to promising directions for future research, including experimental studies [22], interdisciplinary investigations, field research, and economic and judicial management analyses.

In summary, this study contributes to advancing the understanding of how technology can enhance the credibility of judicial hearings via videoconference, offering concrete proposals grounded in in practical judicial experience and in technical and legal knowledge. The gradual and judicious implementation of the suggested functionalities, accompanied by continuous evaluation and evidence-based adjustments, has the potential to significantly transform the quality of judicial service in digital environments [2], expanding access to justice without compromising procedural reliability.

The future of the Brazilian judicial system, like many others globally, will inevitably be hybrid, integrating in-person and remote components tailored to the specific needs of each case and context. In this scenario, the development of technological tools specifically designed for the judicial environment, such as those proposed in this study, represents not only a response to immediate challenges but a strategic path toward a more sustainable, equitable, and resilient justice system.

## Glossary of Technical Terms

To facilitate understanding of the specific technical concepts addressed in this article, we present the definitions of the main terms used:

**Automation bias**: Cognitive tendency to place excessive trust in decisions or suggestions provided by automated systems, which may reduce critical evaluation and lead to uncritical acceptance of outputs, even when they conflict with other available evidence or human judgment.

**Behavioral analysis**: Set of technologies that interpret human actions and expressions—such as eye movements, head orientation, facial tension, and other non-verbal cues—to identify patterns potentially related to attention, stress, or deception, which may be relevant in credibility assessments during judicial proceedings.

**Breakout rooms**: Simultaneous rooms that allow dividing videoconference participants into smaller, isolated groups, enabling private conversations (such as between lawyers and clients) within the same videoconference session.

**Eye contact (Simulated visual contact)**: Technology that adjusts the apparent direction of the participant's gaze to create the illusion of direct eye contact with other videoconference participants, even when the person is looking at different areas of the screen.

**Eye tracking**: Technology that uses the device's camera to map and analyze the direction, fixation, and movements of the user's eyes, allowing identification of where the person is looking and detecting suspicious patterns of visual deviation.

**LGPD (Lei Geral de Proteção de Dados)**: General Data Protection Law. Brazilian legislation (Law No. 13,709/2018) that regulates the processing of personal data by natural or legal persons, public or private, establishing principles, rights, and obligations aimed at protecting privacy and information security. The LGPD, Brazil's equivalent to the EU's General Data Protection Regulation (GDPR), imposes strict requirements on processing personal data, including biometric information used in monitoring technologies.

**Mosaic mode**: Visual arrangement in videoconferences where all participants are displayed simultaneously in tiles (or "windows") of similar size, allowing collective visualization and facilitating the reading of reactions and non-verbal interactions in a group.

**Proctoring (Remote supervision)**: Monitoring system applied in online assessments that uses technologies such as eye tracking, presence detection, and behavioral analysis to ensure the reliability and authenticity of participation, preventing fraud and unauthorized consultations.

**Sandbox (Isolated environment)**: Segregated computational execution space from the main operating system, which prevents applications from accessing unauthorized system resources or other running programs, ensuring greater security and control.